\documentclass[10pt]{elsarticle}
\usepackage{graphicx}
\usepackage{amsmath,amssymb,amsthm,bm}
\usepackage{gensymb}
\usepackage{cancel}
\usepackage{float}
\usepackage{miller}
\usepackage{subcaption}
\usepackage[margin=1in]{geometry}
\usepackage{esint}

\biboptions{sort&compress}
\begin{document}

\begin{frontmatter}

\title{Understanding the anomalous thermal behavior of $\Sigma3$ grain boundaries in a variety of FCC metals}
\author[rvt]{Ian Chesser\corref{cor1}}
\ead{ichesser@andrew.cmu.edu}
\author[rvt]{Elizabeth Holm}
\ead{eaholm@andrew.cmu.edu}
\cortext[cor1]{Corresponding author}

\address[rvt]{Department of Materials Science and Engineering,
Carnegie Mellon University,
5000 Forbes Avenue Wean Hall 3325, Pittsburgh, PA 15213}

\begin{abstract}
We present a case study of the complex temperature dependence of grain boundary mobility. The same general incoherent twin boundary in different FCC metals is found to display antithermal, thermal, and mixed mobility during molecular dynamics synthetic driving force simulations. A recently developed energy metric known as the generalized interfacial fault energy (GIFE) surface is used to show that twin boundaries moving in an antithermal manner have a lower energetic barrier to motion than twin boundaries moving in a thermally activated manner. Predicting the temperature dependence of grain boundary motion with GIFE curves stands to accelerate research in grain boundary science. 
\end{abstract}
\end{frontmatter}


\section*{Introduction}
Predicting the rate at which a grain boundary moves under an applied driving force is a complex problem eminently relevant to the design of ceramic and metallic microstructures. Grain boundary mobility is defined via linearized reaction rate theory as the proportionality constant between an applied driving force and resultant grain boundary velocity \cite{Gottstein2009,Upmanyu2007}:
\begin{align}
	v = M P
\end{align}
where $P$ is an applied driving force with units of energy density or equivalently pressure, $v$ is grain boundary velocity, and $M$ is mobility. 
Grain boundary motion is often considered to have an ideal Arrhenius temperature dependence on mobility  \cite{Cantwell2015}: 
\begin{align}
	M = M_o\,\exp\Big(\frac{-E_a}{k_b T}\Big)
\end{align}
$M$ is the steady state mobility, $M_o$ is a prefactor related to the attempt frequency of grain boundary motion, $E_a$ is an activation energy that represents the energetic barrier to grain boundary motion, and $k_b\,T$ is the thermal energy of the system. The diverse temperature dependence of grain boundary mobility in simulations and experiments imply that Eqn $(2)$ is a vast oversimplification \cite{Sutton1996,Balluffi2005,Gottstein2009,Cantwell2015,Olmsted2009,Homer2014,Ulomek2016,Race2015a}. 

Molecular dynamics (MD) simulations have shown that grain boundaries frequently deviate from an ideal Arrhenius temperature dependence of mobility. Eighty-nine grain boundaries in the 388 boundary mobility survey by Olmsted et. al showed antithermal mobility over some temperature range \cite{Olmsted2009,Holm2010}. Antithermal interface motion is characterized by a decrease in interface mobility with increasing temperature. It is speculated to be important to \textit{cold fast} processes such as room temperature grain growth in nanocrystalline metals,  and is experimentally observed in \textit{hot slow} grain growth in ceramic materials like SrTiO$_{3}$ \cite{Cantwell2015,Cantwell2014,Obrien2016}. Understanding antithermal grain boundary motion is related to many other open problems in interface mobility, including stress driven grain growth and rotation at cryogenic temperatures, twinning and de-twinning processes, and twin defect interactions \cite{Zhang2017,Wang2010,Li2013,Li2011a,Wang2011,Fan2017}.

$\Sigma3$ twin-related boundaries comprise over half of the antithermal boundaries in the Olmsted survey \cite{Olmsted2009}. Remarkably, twin boundary mobility spans several orders of magnitude in MD simulations and shows antithermal, thermal, or mixed dependence on temperature as a function of boundary plane crystallography \cite{Priedeman2017}. %A small number of \textit{mixed mobility} boundaries show thermally activated behavior at low temperatures and antithermal behavior at high temperatures \cite{Priedeman2017}. 
Previous MD mobility simulations have only considered twin boundary mobility in Ni . In this study, we consider the mobility of the same general twin boundary in twelve different FCC metals, testing fifteen interatomic potentials in total, four for Ni. 

\section*{Methods}
We examine the motion of a general $\Sigma3$ \hkl[1 1 1] 60$\degree$ \hkl(11 8 5) grain boundary. The boundary facets into segments of coherent twin (CT) and incoherent twin (ICT) boundary upon energy minimization, consistent with existing faceting models for $\Sigma3$ boundaries \cite{Banadaki2016}. The grain boundary moves via the glide of the incoherent \hkl{1 1 0} facets, as shown in Figure 1(b). The grain boundary has mixed tilt and twist character, and is most simply viewed as a rotation from the CT with \hkl{111} boundary planes by 17 $\degree$ about a \hkl<1 1 2> axis. 
%Faceting into low energy CT and ICT segments is consistent with existing faceting models for $\Sigma3$ boundaries \cite{Banadaki2016}.

\begin{figure}[H]
  \centering
      \includegraphics[width=\linewidth]{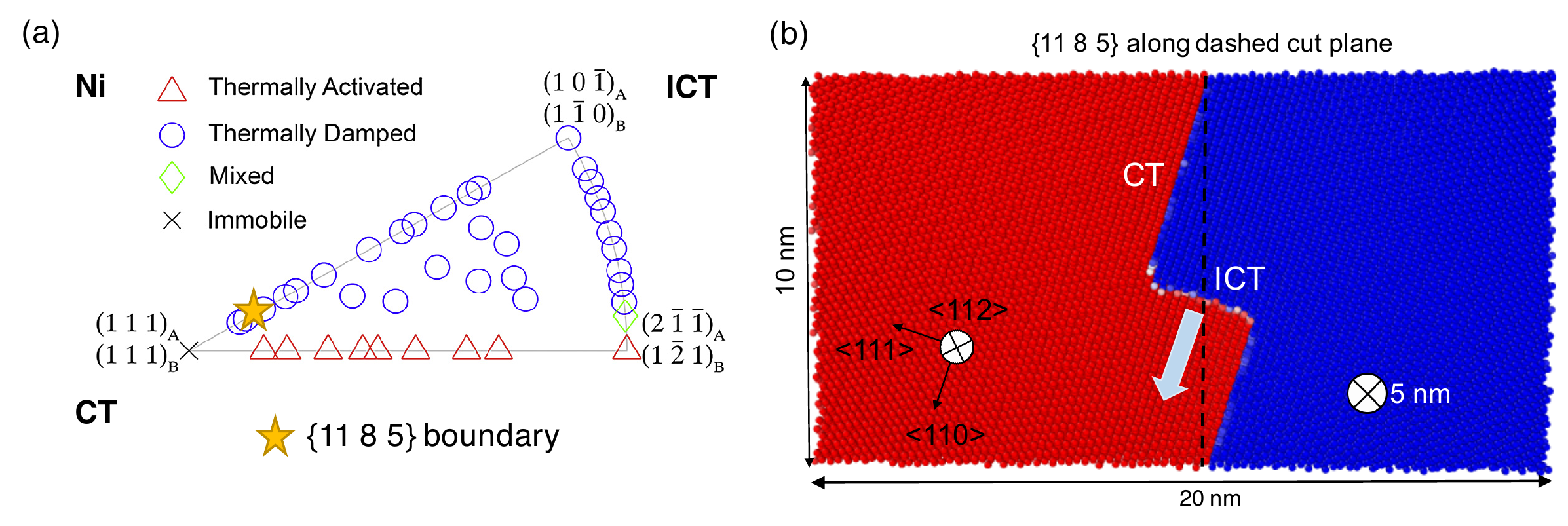}
   \caption{
   (a) Boundary plane fundamental zone with points colored by mobility type \cite{Priedeman2017}. The \hkl{11 8 5} boundary in this work is shown with a yellow star. (b) Faceted \hkl{11 8 5} grain boundary, schematically showing the growth of the blue grain under an energy jump driving force 
  }\label{fig:ex4}
\end{figure}

Twin boundary mobility is examined with the synthetic driving force (SDF) molecular dynamics method \cite{Janssens2006, Ulomek2015} for twelve different embedded atom method (EAM) potentials fit by Sheng et. al \cite{Sheng2011}. 
The Sheng potentials were fit to a variety of elemental properties, including lattice dynamics, mechanical and thermal properties, and the energetics of competing crystal structures, defects, deformation paths, and liquid structures. Three Ni potentials are considered other than the Sheng potential, including the Foiles-Hoyt Ni potential used in prior studies of antithermal grain boundary motion, the Mishin Farkas Ni potential, and the Mendelev Ni potential \cite{Sheng2011,Foiles2006,Mendelev2012,Mishin1999}. We use the following shorthand for Ni potentials in this work: Ni1 = Sheng, Ni2 = Foiles-Hoyt, Ni3 = Mendelev, Ni4 = Mishin/Farkas. 

Grain boundaries are constructed in a simulation box in LAMMPS with periodic boundary conditions, and are minimized via a standard conjugate gradient minimization at 0 K \cite{plimpton1995fast}. All SDF simulations are run in an NPT ensemble with periodic boundary conditions along the grain boundary plane and free surface boundary conditions normal to the cut plane. Temperature is set with a Nose Hoover thermostat and pressures are maintained near zero along the grain boundary plane with a Parrinello Rahman barostat.  A timestep of 2 fs is used to anneal grain boundaries for 0.4 ns before the synthetic driving force is turned on for 1 ns. The SDF method imposes a driving force for boundary motion in a bi-crystal by lowering the free energy of one grain and raising the free energy of the other, causing the low energy grain to grow at the expense of the high energy grain. Boundary position is tracked via the change in order parameter of a fixed region encompassing the shrinking grain. Velocity is computed as the slope of the position-time curve via the bootstrap resampling technique of Race et. al with a smoothing window of 5 ps and a sample window of 20 ps \cite{Race2015}. Finally, mobility is calculated in the linear approximation as velocity divided by the applied driving force. 

For mobility results for different FCC metals, a driving force of 10 meV/atom was used. Previous simulations have found 1 meV/atom to be an optimal driving force for studying the motion of anti-thermal twin boundaries in Ni, since mobility values at 1 meV match values calculated in the zero driving force limit from the random walk method \cite{Trautt2006}. We choose a larger driving force because of the need to generate as much motion as possible in slow moving thermal boundaries over a 1 ns time span. We compare driven motion of Ni potentials using a SDF of 1meV/atom, consistent with previous studies. Note that system size is relevant to energy and mobility convergence because defects can interact across periodic boundaries and modify minimum energy structures and grain boundary dynamics \cite{Humberson2016, Race2014}. A full justification of simulation box lengths with respect to energy and mobility can be found in the work of Humberson and Holm \cite{Humberson2016,Humberson2017,Humberson2017a}. 
%The simulation box for each grain boundary is constructed in LAMMPS with a coincident site lattice (CSL) unit cell with orthogonal basis vectors along the directions shown in Figure 1b. CSL basis vector lengths are calculated in the open source software package GBpy \cite{Banadaki-gbpy}. 10 CSL repeats are chosen along the tilt axis in the \hkl<1 1 2> direction, 5 repeats in the in plane \hkl<6 2 10> direction, and 30 repeats in the  \hkl<11 8 5> direction normal to the ideal grain boundary plane. A full justification of these parameters can be found in the work of Humberson and Holm \cite{Humberson2016,Humberson2017,Humberson2017a}. 

\section*{Results}

The mobility behavior of the same general twin boundary is found to vary with FCC metal, both across EAM potentials for different FCC metals and within EAM potentials for Ni. Figure 2a plots mobility of the $\Sigma3$ \hkl[1 1 1] 60$\degree$ \hkl(11 8 5) boundary for a variety of FCC metals with different stacking fault energies. Mobility is plotted at three homologous temperatures $T_H$ for each metal for a synthetic driving force of 10 meV/atom. The colored background of Figure 2a shows the clustering of FCC metals by mobility type as a function of stacking fault energy. Ca, Sr, Au, Ag, and Cu show antithermal behavior (red) while Pb, Ce, Al, Pt, and Rh show thermal behavior (blue). Pd displays mixed behavior (purple), with the intermediate temperature corresponding to the highest mobility. %Transitions from antithermal to thermal mobility type occur between Sr and Pb, and Cu and Al, while transitions from thermal to antithermal mobility type occur between Ce and Ag. 
Although stacking fault energy is a convenient parametrization of these metals and transitions, it is not a predictor of mobility type. 

Boundary mobility also varies significantly among Ni potentials, as shown in Figure 2b. Ni4 exhibits a weakly thermal trend, while other Ni potentials show overall antithermal trends. Ni2 and Ni4, which have the same measured stacking fault energy, show different mobility type. The goal of the remainder of this work is to rationalize differences in mobility type for the same twin boundary in different FCC EAM potentials in the context of grain boundary structure and energetics. 

\begin{figure}[H]
  \centering
       \includegraphics[scale = 0.6]{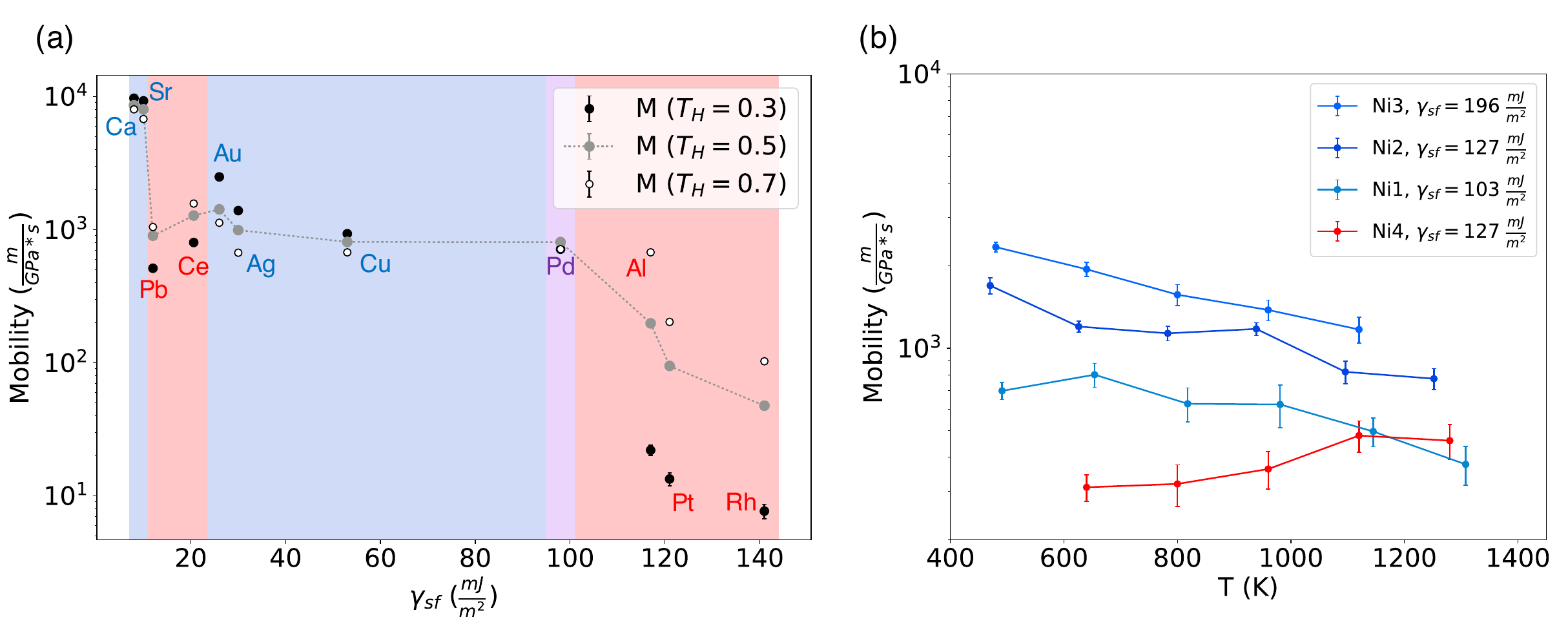}
   \caption{
   (a) Variation in the mobility of the \hkl{11 8 5} boundary with stacking fault energy for different Sheng potentials at three homologous temperatures $T_H$ and a synthetic driving force of 10 meV/atom. Mobility type given by colored background, with red = thermal, blue = antithermal, and purple = mixed. (b) Variation in mobility of \hkl{11 8 5} boundary with temperature for different Ni potentials under a synthetic driving force of 1 meV/atom 
  }\label{fig:ex1}
\end{figure}

The initial ICT facet length structure of the \hkl{11 8 5} boundary varies with interatomic potential. Representative facet structures are shown in Figure 3 after a 0.4 ns aneal and a small amount of motion under a SDF. In Cu, a single facet is present with height $15\,a_{\hkl(111)}$, referred to here as h(15) following the convention of Hirth \cite{Hirth2016}. In Pb and Al, multiple facets are present after annealing. From left to right, Pb has a facet height distribution of $\{h(6),h(9)\}$, while Al has a facet height distribution of $\{h(3),h(9),h(3)\}$. %In Figure 3, the grain boundary plane is outlined in dark red, while triplets of \hkl{111} planes are highlighted in red, blue, and white according to in plane displacements of atoms in a \hkl<1 1 2> direction along \hkl{1 1 1} planes. 

\begin{figure}[H]
  \centering
      \includegraphics[width=\linewidth]{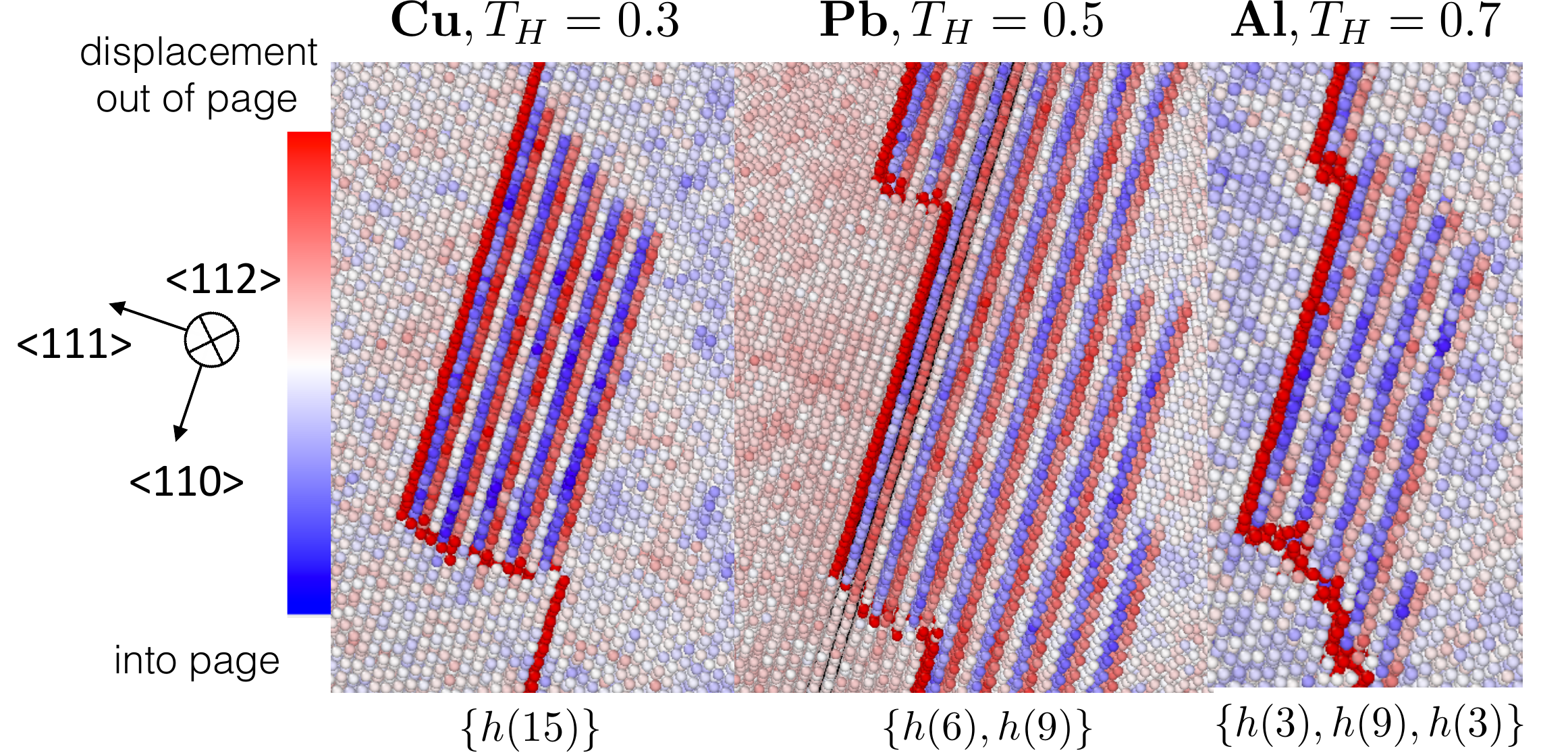}
   \caption{
   Finite facet structure of the annealed \hkl{11 8 5} boundary in Cu, Pb, and Al after a small amount of boundary motion under a SDF of 10 meV/atom at various homologous temperatures. The grain boundary plane is 	                     outlined in dark red, while triplets of \hkl{111}  planes are highlighted in red, blue, and white according to in plane displacements of atoms in a \hkl<112> direction. The h(3) facet unit is clearly defined by white/blue/red triplets.
  }\label{fig:ex2}
\end{figure}

The h(3) facet is the fundamental motion unit during \hkl{11 8 5} boundary motion. h(3) facets can coalesce during boundary motion, presumably to reduce corner energy at facet junctions \cite{Daruka2004, Wu2009, Medlin2017} . All elements that show anti-thermal motion show a single facet after annealing. Although most thermal boundaries retain multiple facets, Pb, Ce, and Ni4 have facets of height h(15) after annealing yet move in a thermal manner. Facet coalescence is a necessary but insufficient condition for anti-thermal motion in these simulations.
%\begin{enumerate}
%  \item Facet heights are always integer multiples of h(3) $= 3\,a_{\hkl(111)}$
%  \item Maximum facet height is set by the simulation box length in the \hkl<6 2 10> direction, which determines the geometrically necessary length of incoherent twin boundary. For the boundaries considered, $\Sigma h_i = $ h(15)
%  \item At 0 K, the \hkl{11 8 5} boundary consists of five separated h(3) facets. These facets coalesce at finite temperature, presumably to reduce corner energy at facet junctions. \cite{Daruka2004, Wu2009, Medlin2017} 
%  \item A given facet distribution may coalesce to a single facet with increased annealing time. Upon further annealing of the Pb boundary at $T_h = 0.3$, for example, the two smaller facets in Fig 5 are observed to coalesce to a single larger facet. Al, however, maintains multiple facets throughout its motion at all temperatures considered. 
%\end{enumerate}

%Both Ni2 and Ni4 have the same stacking fault energy and faceted configuration, but differ in mobility type. Similarly, Pb and Ce have facets of height h(15) after annealing, similar to anti-thermal boundaries, yet move in an apparently thermal manner. These observations motivate us to examine the disconnection structure and motion mechanism of ICT facets in order to discern differences relevant to mobility type. Ultimately, we construct an energy metric that directly measures the energy barrier to ICT facet motion, but start by considering the generalized stacking fault and twin fault energy surface for the potentials considered.

Since facet structure does not distinguish mobility type, we instead consider energy metrics relevant to twin related boundaries. A twinned crystal can be created by applying successive shearing operations to a perfect crystal. Starting with a perfect crystal block, a stacking fault is generated by applying a rigid translation of the upper half of the block along a \hkl{111} plane in the \hkl<112> direction. Performing such a translation incrementally and measuring energy after each shift leads to the generation of a stacking fault energy curve. The generalized stacking fault energy surface (GSFE) includes shifts in the perpendicular in plane \hkl<110> direction as well as the \hkl<112> direction \cite{Vitek1968}. A single layer of twinned crystal can be created by starting with a stacking fault, and rigidly translating all atoms on top of the stacking fault by $a/6$\hkl<112>. The energy surface corresponding to this procedure is known as the generalized twin fault energy surface (GTFE) \cite{VanSwygenhoven2004}. The GTFE is analogous in concept to the GSFE, but has a different starting configuration. GSFE and GTFE surfaces were measured for each interatomic potential (not shown) with tabulation of stable and unstable stacking and twinning fault energies. Overall, no single energy barrier from the GSFE or GTFE surface was found to cluster elements by mobility type in a way that added new information compared to classification by stacking fault energy. We therefore proceed to construct an energy metric that captures more complexity of ICT facet motion. 

%Starting with a perfect crystal block, a rigid translation is applied to the upper half of the block along a \hkl{111} plane in the \hkl<112> direction. A translation of $a/6$\hkl<112> leads to a stacking fault, an interruption in the FCC stacking sequence. Performing such a translation incrementally and measuring energy after each shift leads to the generation of a stacking fault energy curve. The generalized stacking fault energy surface (GSFE) includes shifts in the perpendicular in plane \hkl<110> direction as well as the \hkl<112> direction \cite{Vitek1968}. Energies are calculated via a conjugate gradient energy minimization that only considers relaxation normal to the interface. If relaxations were considered along the plane of the fault, the system would displace to the minimum energy states of either a perfect crystal or stacking fault, and no energy barriers would be constructed. Starting with a stacking fault, rigid translations of all atoms on top of the stacking fault by $a/6$\hkl<112> lead to a single layer of twinned crystal. An energy surface can be created for this shifting procedure known as the generalized twin fault energy surface (GTFE) \cite{VanSwygenhoven2004}. The GTFE is analogous in concept to the GSFE but has a different starting configuration. The growth of a new twin grain can be imagined as repeated shifts by $a/6$\hkl<112> on subsequent \hkl{111} planes. 

Translation of ICT facets during boundary motion is accomplished by correlated shuffling of atoms along triplets of \hkl{111} planes, or h(3) units.  Each h(3) facet shows a characteristic displacement pattern as illustrated in Figure 3. Atoms in adjacent \hkl(1 1 1) planes are observed to shuffle along shockley partial vectors s(1) = a/6\hkl[1 -2 1] and s(2) = a/6\hkl[-2 1 1], colored in blue and red in Figure 3. Every third plane does not undergo displacement, as shown in white. It is observed that atoms two to three layers in front of the ICT facet shuffle incrementally into the new grain orientation swept out by the ICT facet. Since h(3) facets do not shear the lattice upon motion, they can be considered disconnections of pure step character \cite{han2018grain}.  Motion of collections of h(3) facets in the \hkl{11 8 5} boundary is a type of step flow motion. Step flow motion of nano-facets has been observed in other general grain boundaries \cite{Hadian-facet}. We proceed to relate the energetics of step flow motion of the \hkl{11 8 5} boundary to mobility type for a variety of FCC metals. 

An energy metric known as the generalized interface fault energy (GIFE) surface that calculates a minimum energy path for the motion of a disconnection has recently been developed by Barrett et. al  \cite{Barrett2017}. We reimplement the general method here in the context of the h(3) \hkl{110} ICT facet, the fundamental motion unit of the \hkl{11 8 5} boundary. The GIFE technique consists of the following steps: 
\begin{enumerate}
   \item An initial and final state are created before and after the passage of a disconnection of known character. We construct a single h(3) facet in an initial and final minimized configuration separated by $a/2$\hkl<110>. To isolate the motion of a single facet, we consider two free surface boundary conditions parallel to the CT and ICT facets and periodic boundary conditions in the \hkl<112> direction (into the page in Figure 4a). We choose a system size such that the initial and final energies of the facets are the same, indicating that the effect of free surfaces are negligible. 
   \item A displacement map is calculated between nearest neighbors in the initial and final state. This displacement map captures relevant shuffling vectors
   \item The displacement map is incrementally applied to the initial state, with energy calculated after each increment. Like the GSFE and GTFE, relaxation is only considered normal to the interface plane. If relaxation is not restricted in this way, the system will return to the initial or final state as its preferred state. 
\end{enumerate}

GIFE results are shown for the motion of a h(3) facet in a variety of FCC metals in figure 4(b). The GIFE curve separates thermally activated boundaries from anti-thermal boundaries by energy barrier. The elements with thermally activated \hkl{11 8 5} boundaries: Rh, Al, Pt, Ce, and Pb, have a higher energy barrier to h(3) facet motion than elements with antithermal boundaries:  Au, Cu, Ag, Ca, and Sr. The GIFE barrier for the mixed-mode Pd falls between the two classes. Note that the GIFE barrier is calculated as an excess energy per area relative to the initial faceted state, which varies for different elements. The magnitude of the GIFE barriers of the antithermal elements in figure 4(b) spans 0.01 to 13 $mJ/m^2$, while barriers for thermal elements fall in the range 24-204 $mJ/m^2$. 
%(Provide value in eV/atom here, using length scale of transformation normal to interface)

GIFE curves for the four Ni potentials are shown in figure 4(c). The barrier for Ni1, classified as an antithermal boundary, is anomalously high at around 70 $mJ/m^2$ and falls above the barriers for thermal Ce and Pb in figure 4(b). GIFE barriers are sensitive to small modifications in shuffling path which arise from changes in the initial and final minimized configuration.  We demonstrate this sensitivity using the four Ni potentials as a case study. All Ni potentials are trapped in the initial and final state assumed by Ni2 such that the Ni1,Ni3, and Ni4 initial states are metastable. The GIFE barriers calculated for metastable Ni1,Ni3, and Ni4 are shown as dashed lines in figure 4(c). The GIFE barrier for Ni4 increases by around 20 $mJ/m^2$, while the barrier for Ni1 decreases by 60 $mJ/m^2$ to a value consistent with the antithermal elements in figure 4(b). Trapping the Ni potentials in the minimized states of Ni1, Ni3, and Ni4 lead to similar reordering effects. In three out of four cases, the thermal boundary corresponding to Ni4 has the highest energy barrier. The sensitivity of GIFE barriers to initial minimized configuration points to the need for more systematic convergence of GIFE barriers with respect to initial conditions. Like grain boundary energy, GIFE barriers should be converged with respect to microscopic degrees of freedom. Interestingly, the boundaries with lowest GIFE barrier may correspond to metastable initial states. 

%The GIFE method is a powerful technique because it captures degrees of freedom relevant to the motion mechanism of the bicrystal that are not included in the dichromatic pattern, namely, displacement fields imposed by relaxation of the interface and corresponding modifications to shuffling path. Such relaxations are observed in TEM studies of real twin boundaries \cite{Pond1977,Lucadamo2017}. The GIFE method is similar to the nudged elastic band (NEB) method, but sacrifices some accuracy for computational efficiency \cite{legros-kink}. The GIFE method linearly interpolates a path from an initial to final point set representing two grain orientations and extracts an energy barrier. The NEB method allows this path to curve in order to reduce the energy barrier further. 

The same general twin boundary geometry shows remarkable variation in mobility type with EAM potential under a synthetic driving force. We have found a correspondence between GIFE energy barriers to grain boundary motion and grain boundary mobility type: antithermal boundaries have low GIFE barriers to motion relative to thermal boundaries. Out of the energy metrics considered, only the GIFE barrier was able to separate elements by mobility type class. Small changes in boundary structure can significantly affect the energy barrier given by the GIFE method, highlighting the need to develop more systematic ways to select and validate initial states. The current results reaffirm the fact that semi-empirical EAM potentials, even those fit to a wide range of defect energies, may not accurately reproduce complex kinetic phenomena like grain boundary mobility. DFT calculations of GIFE curves should be used as a way to standardize EAM potentials with respect to mobility type. The combination of DFT calculations and a broader survey of GIFE barriers for different grain boundaries stands to greatly improve our understanding of the thermal behavior of grain boundaries. 

\begin{figure}[H]
  \centering
      \includegraphics[width=\linewidth]{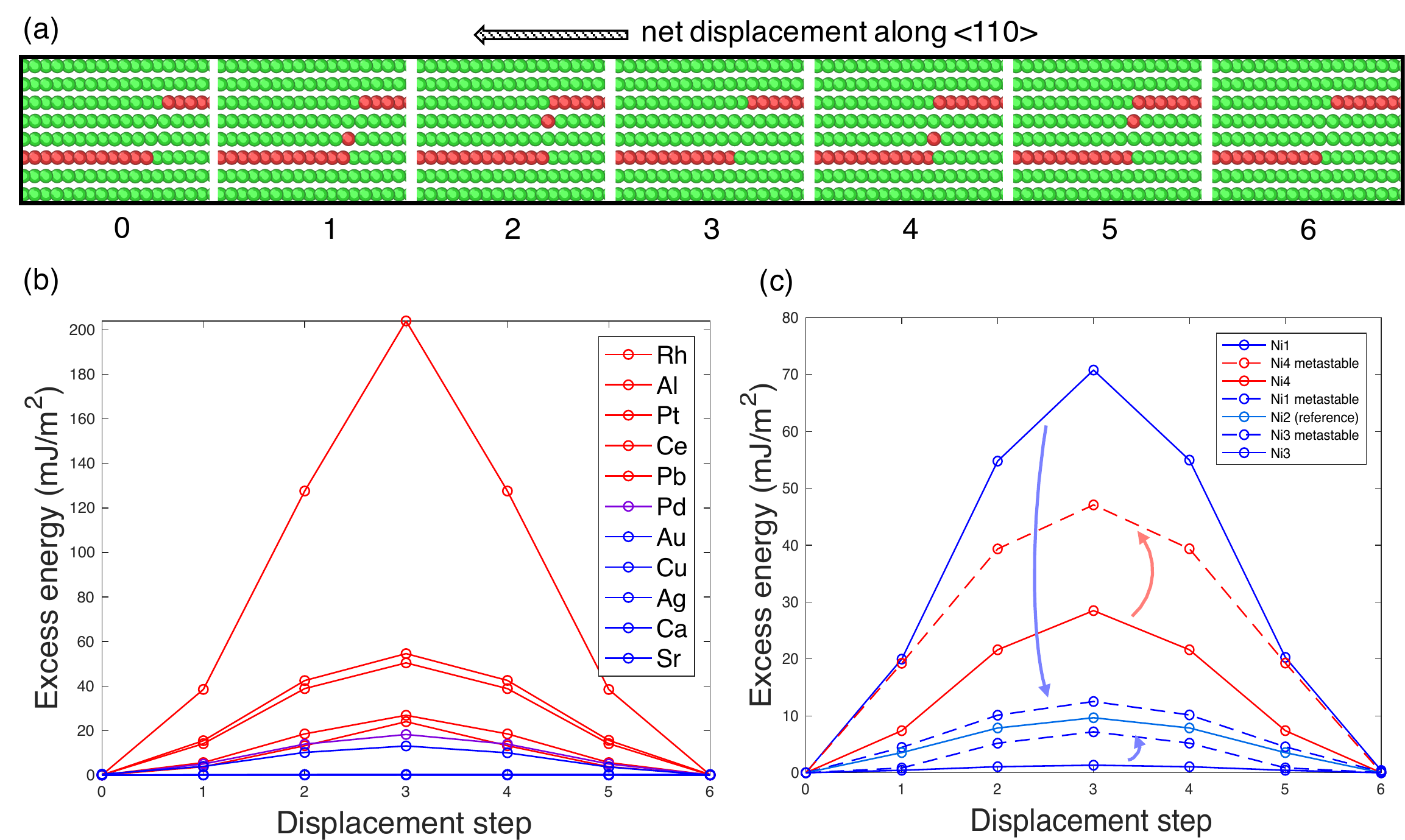}
   \caption{
   (a) GIFE process: a single \hkl{110} facet incrementally moves from an initial to final state of equivalent energy. (b) GIFE results for h(3) \hkl{110} ICT facet motion in a variety of FCC metals. Elements are ordered in legend from highest to lowest barrier and colored by mobility type, with red = thermal, blue = antithermal, purple = mixed. Low energy barriers correspond to antithermal motion. (c) GIFE results for Ni potentials (solid lines) and Ni potentials trapped in the minimized configuration of Ni2 (dashed). Arrows denote change in energy from the minimum energy to metastable trajectories. 
  }\label{fig:ex4}
\end{figure}

\section*{Acknowledgements}
This work was supported by the National Science Foundation GRFP Grant No. 1252522 and award number DMR-1710186, as well as the W.M. Keck Foundation. Thanks to Jonathan Humberson, David Srolovitz, Jian Han, Howard Sheng, Yuri Mishin, G\"{u}nter Gottstein, Xiaoting Zhong, Greg Rohrer, Marc De Graef, David Laughlin, Tony Rollett, and Christopher Barrett for helpful discussions regarding the work.
%\section*{References}
%\bibliographystyle{abbrvnat}
\bibliographystyle{unsrtnat}
\bibliography{ms}

\end{document}